\documentclass[conference]{IEEEtran}
\IEEEoverridecommandlockouts
\usepackage{cite}
\usepackage{amsmath,amssymb,amsfonts}
\usepackage{algorithmic}
\usepackage{graphicx}
\usepackage{textcomp}
\usepackage{xcolor}
\usepackage{multirow}
\usepackage{comment}

\def\BibTeX{{\rm B\kern-.05em{\sc i\kern-.025em b}\kern-.08em
    T\kern-.1667em\lower.7ex\hbox{E}\kern-.125emX}}
\begin{document}

\title{Noninvasive ultrasound for Lithium-ion batteries state estimation\\

}
\begin{comment}
\author{\IEEEauthorblockN{1\textsuperscript{st} Simon Montoya-Bedoya}
\IEEEauthorblockA{\textit{Department of Nanotechnology Engineering} \\
\textit{Universidad Pontificia Bolivariana}\\
Medellin, Colombia \\
simon.montoya@upb.edu.co}
\and
\IEEEauthorblockN{2\textsuperscript{nd} Miguel Bernal}
\IEEEauthorblockA{\textit{Verasonics SAS}\\
Medellin, Colombia  \\
miguelbernal@verasonics.com}
\and
\IEEEauthorblockN{3\textsuperscript{rd} Laura A. Sabogal-Moncada}
\IEEEauthorblockA{\textit{Department of Nanotechnology Engineering} \\
\textit{Universidad Pontificia Bolivariana}\\
Medellin, Colombia \\
laura.sabogal@upb.edu.co}
\and
\IEEEauthorblockN{4\textsuperscript{th} Hader V. Martinez-Tejada}
\IEEEauthorblockA{\textit{Department of Mechanical Engineering} \\
\textit{Universidad Pontificia Bolivariana}\\
Medellin, Colombia  \\
hader.martinez@upb.edu.co}
\and
\IEEEauthorblockN{5\textsuperscript{th} Esteban Garcia-Tamayo}
\IEEEauthorblockA{\textit{Department of Nanotechnology Engineering} \\
\textit{Universidad Pontificia Bolivariana}\\
Medellin, Colombia  \\
esteban.garciat@upb.edu.co}
}
\end{comment}

\author{\IEEEauthorblockN{Simon Montoya-Bedoya\IEEEauthorrefmark{1},
Miguel Bernal\IEEEauthorrefmark{2},
Laura A. Sabogal-Moncada\IEEEauthorrefmark{1}, 
Hader V. Martinez-Tejada\IEEEauthorrefmark{3},\\
Esteban Garcia-Tamayo\IEEEauthorrefmark{1}}

\IEEEauthorblockA{\IEEEauthorrefmark{1}Department of Nanotechnology Engineering, Universidad Pontificia Bolivariana, Medellín, Colombia}
\IEEEauthorblockA{\IEEEauthorrefmark{2}Verasonics SAS, Medellín, Colombia}
\IEEEauthorblockA{\IEEEauthorrefmark{3}Department of Mechanical Engineering, Universidad Pontificia Bolivariana, Medellín, Colombia}
% <-this % stops an unwanted space
}

\maketitle

\begin{abstract}
Lithium-ion battery degradation estimation using fast and noninvasive techniques is a crucial issue in the circular economy framework of this technology. Currently, most of the approaches used to establish the battery- state (i.e., State of Charge (SoC), State of Health (SoH)) require time-consuming processes. In the present preliminary study, an ultrasound array was used to assess the influence of the SoC and SoH on the variations in the time of flight (TOF) and the speed of sound (SOS) of the ultrasound wave inside the batteries.  Nine aged 18650 Lithium-ion batteries were imaged at 100\% and 0\% SoC using a Vantage-256 system (Verasonics, Inc.) equipped with a 64-element ultrasound array and a center frequency of 5 MHz (Imasonic SAS). It was found that second-life batteries have a complex ultrasound response due to the presence of many degradation pathways and, thus, making it harder to analyze the ultrasound measurements. Although further analysis must be done to elucidate a clear correlation between changes in the ultrasound wave properties and the battery state estimation, this approach seems very promising for future nondestructive evaluation of second-life batteries

\end{abstract}


\section{Introduction}

Electrochemical energy storage in form of Lithium-ion batteries represents the most efficient tool for the decarbonization and electrification process due to three main factors:
(i) increase the use of renewable energy, (ii) reduce the energy dependence on oil and gas and, (iii) promote the circular economy by repurposing batteries that reached the end of their first useful life to other applications called second-life batteries (SLBs) \cite{b1}. Likewise, the versatility of their architecture (pouch, cylindrical or prismatic) allows them to be used in portable and stationary energy applications.

Even more, sales of electric vehicles (EVs) have been growing all over the world \cite{b2}, and considering that the batteries of these systems are in some cases still being changed at 70-80\% of their nominal capacity \cite{b3}, the use of second-life batteries (SLBs) is creating an environment conducive to a circular economy (new) market in the world \cite{b4}. In addition, from the point of view of the use of components and materials, it is also the promoter of the Recycled batteries (RBs) -i.e. batteries without an opportunity to use –. In the case of two-wheeled vehicles (motorcycles and bicycles), these are still a very young market, for which there is not even a global regulation coordinated between the main markets (i.e. EU, USA, China). In this regard, it is to be expected that in the next decade the return flows of SLBs and RBs, will have an increasing commercial interest as the lack of sources of critical materials, especially cobalt, nickel, and manganese will become increasingly evident. Therefore, from the perspective of a circular economy and considering what would be the use of SLBs and RBs, the future scenario for the most representative electric vehicle segments (cars, campers, vans, pickups, and two-wheelers) would also mean energy savings, GHG reduction, and lower water consumption.

In this sense, accurate estimation methods are required to evaluate the Lithium-ion batteries degradation (which depends on cathode, anode, electrolyte, and separator) and they remaining useful life to guarantee safety, performance, and longevity in their applications \cite{b5}, \cite{b6}. Degradation mechanisms in a macroscopic view result in two principal aspects, a capacity loss and an increase in the internal resistance. To estimate these variables is necessary to know two key metrics that describe the current battery state in terms of its available energy (State of Charge (SoC)) and current capacity (State of Health (SoH)). SoC is an indirect measurement that represents the relation between the average Lithium-ion concentration in the anode surface at a specific moment compared to the maximum Lithium-ion concentration that could be stored in the electrode, and it is usually expressed in percentage points (i.e., 100\% full-charge and 0\% full-discharge). On the other hand, the SoH indicates how much the battery has aged, expressed by the relation between its current and nominal capacity. Its values can go from 0\% to 100\%. Thus, the SoC estimation is one of the most important features of today's battery management systems (BMS) and also due to the support in precise parameters estimation (such as the accuracy of SOH simulation) and to prolong the battery life \cite{b6}, \cite{b7}.

However, Lithium-ion batteries are complex systems, and their behavior depends on the demands of users and factors related to the design (materials and mechanical stress), production (methodology), application (temperature, current load, etc.), and internal reactions \cite{b8}, \cite{b9}. For that reason, previous studies have focused on the development of non-destructive techniques, such as ultrasound, to provide information under operating conditions of the battery and allow an accurate estimation of SoH/SoC \cite{b10}, \cite{b11}. Hsieh et al. (2015) showed a relation between SoC estimations and density distribution in pouch cell and 18650 cylindrical cells \cite{b12}. In their work, they noted that the time of flight (TOF) amplitudes of ultrasonic pulses change with the number of charge/discharge cycles. Later on, Davies et al. (2017) estimated the SoC based on the measure TOF on batteries showing a correlation in the shift of the arrival of the ultrasonics pulses during a full charge-discharge cycle \cite{b11}. \\
The aim of this study was to assess the influence of the SoC and SoH on the variations in the time of flight (TOF) and the speed of sound (SOS) of the ultrasound waves inside the batteries with an ultrasonic array for 18650 second-life Lithium-ion batteries.

\section{Methods}

\subsection{Batteries' conditions}
\begin{figure}[htbp]
\centering
\centerline{\includegraphics[width=9cm]{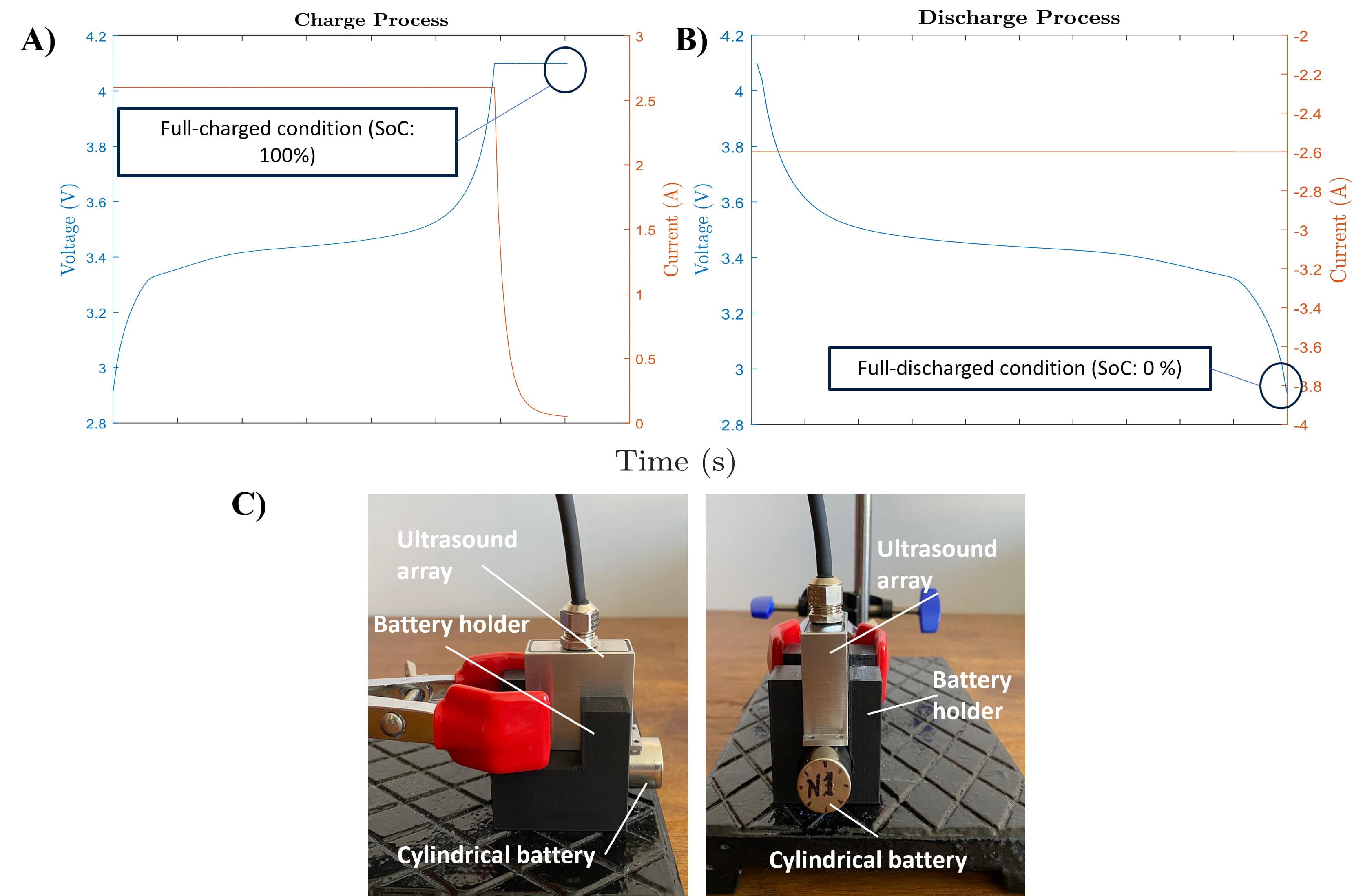}}
\caption{Battery processes: \textbf{A)} charging and \textbf{B)} discharging. \textbf{C)} Ultrasound experimental setup}
\label{fig1}
\end{figure}

Nine aged (second-life) commercial LG 18650 Lithium- ion batteries with a nominal capacity of 2600 mAh and a voltage operation range of 2.75 V (i.e., SoC 0\%) to 4.2 V (i.e., SoC 100\%) were tested in this work. These batteries were obtained from BATx (a Colombian company specialized in second-life battery applications), from an electric two-wheel vehicle that reach its first-life threshold. BATx estimated their SoH using a coulomb counting protocol and classified the batteries into 3 groups based on their SOH (Table \ref{tab1}). A Constant-Current-Constant-Voltage (CCCV) charging protocol was used, which required a first CC charge until a maximum cut-off voltage level (at 4.2 V) followed by a CV stage until the current decrease to 5\% of the nominal current (Fig. \ref{fig1}A). To set batteries in the SOC of 0\% state, the batteries were discharged with a CC method from high to low cut-off voltage levels (Fig. \ref{fig1}B).

\begin{table}[htbp]
\caption{Batteries’ conditions}
\begin{center}
\begin{tabular}{|c|c|c|c|}
\hline
\multicolumn{4}{|c|}{{\color[HTML]{333333} \textbf{Experiments' conditions}}} \\ \hline
Groups                          & Name             & Actual capacity (mAh)       & SoH       \\ \hline
                                & Battery 11       & 2594                        & 1.00      \\  
                                & Battery 12       & 2488                        & 0.96      \\  
\multirow{-3}{*}{Group 1}       & Battery 13       & 2389                        & 0.92      \\ \hline
                                & Battery 21       & 2314                        & 0.89      \\  
                                & Battery 22       & 2246                        & 0.86      \\  
\multirow{-3}{*}{Group 2}       & Battery 23       & 2152                        & 0.83      \\ \hline
                                & Battery 31       & 2063                        & 0.79      \\  
                                & Battery 32       & 2037                        & 0.78      \\  
\multirow{-3}{*}{Group 3}       & Battery 33       & 2020                        & 0.78      \\ \hline
\end{tabular}
\label{tab1}
\end{center}
\end{table}

\subsection{Ultrasound acquisitions}
Batteries were imaged at 100\% (charged) and 0\% SOC (discharged) conditions using an ultrasound array with 64 elements (0.5-mm pitch) and a center frequency of 5 MHz (Imasonic SAS). The ultrasound array was placed longitudinally on the battery, assuring that the imaging plane cut the battery’s diameter (Fig. \ref{fig1}C). Ten ultrasound frames were acquired using a Vantage 256 system (Verasonics, Inc.) with plane-wave imaging sequence at a frame rate of 1 kHz for each battery. The cell was then rotated every 45 degrees (i.e., 8 locations) and the acquisitions were repeated (Fig. \ref{fig1}C).

\subsection{Time of flight and speed of sound measurements}
\begin{figure}[htbp]
\centering
\centerline{\includegraphics[width=9cm]{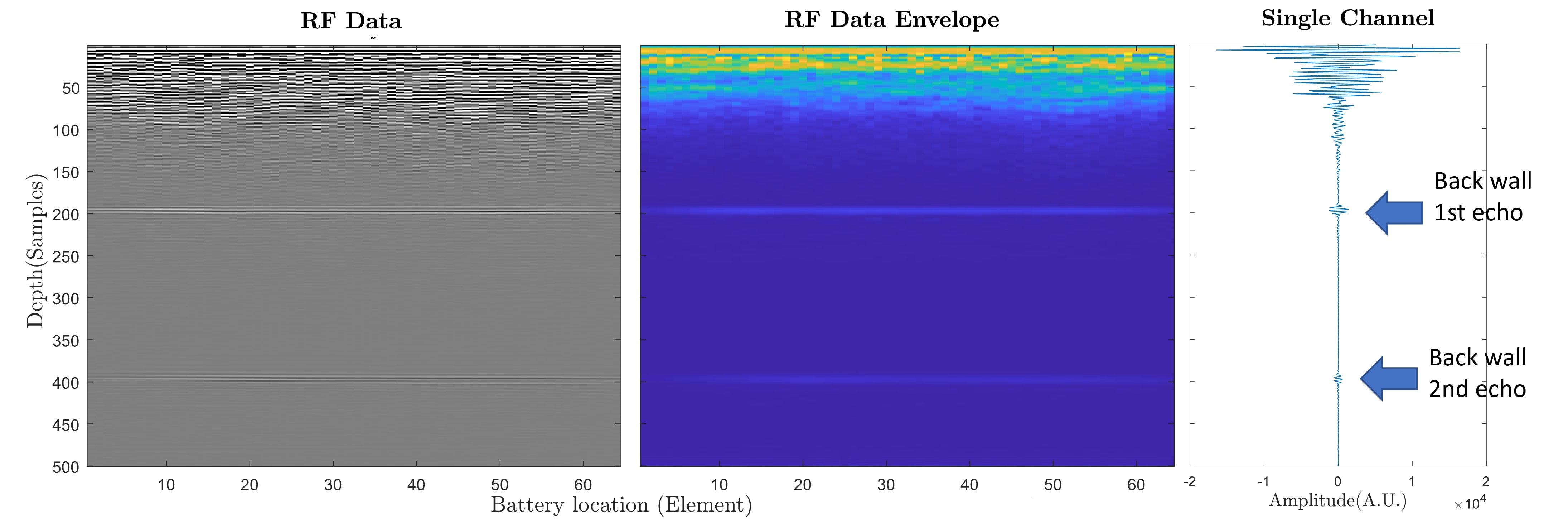}}
\caption{Radio-frequency amplitude signals for each of the 64 channels}
\label{fig2}
\end{figure}

The TOF and SOS were calculated using an autocorrelation function on the backscatter signal from the 1\textsuperscript{st} and 2\textsuperscript{nd} back wall echoes (samples around 200 and 400, respectively in Fig. \ref{fig2}). This process was performed for each acquisition and element in the array. Given the sampling rate of 20 MHz and the peak form the autocorrelation function, the TOF shift was calculated. In addition, given the dimensions of the 18650 batteries of 18-mm diameter and the TOF, the SOS was calculated inside the battery. Furthermore, the data measured from the first and last 6 elements of the probe were discarded to avoid any artefacts in the signal due to coupling between the battery and the probe.

\subsection{'In operando' measurement}
A first ’In operando’ experiment was conducted using the same conditions for a charge/discharge cycle while acquiring continuous ultrasound data every 30 seconds This experiment was carried on battery 13, using an acetate hand-made case to support the battery holder shown in Fig. 1 to maintain electrically isolate the battery contacts, which was submerged in water during the whole process to ensure coupling with the ultrasound transducer. 
\section{Results}
\begin{figure}[htbp]
\centering
\centerline{\includegraphics[width=9cm]{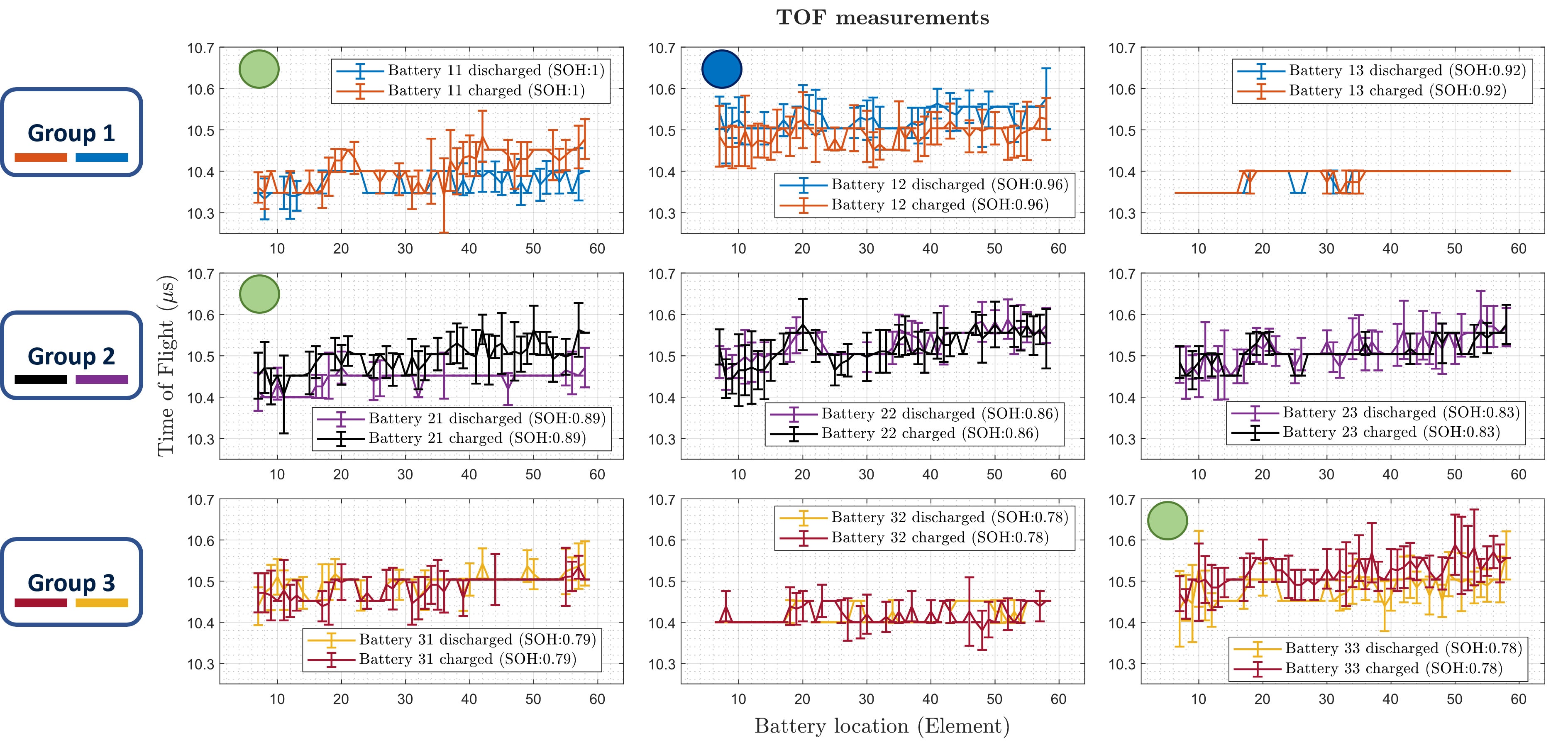}}
\caption{Time of flight estimations for each group and each battery for both tested conditions. (The Green dots indicate the batteries where the TOF was higher for a 100\% SoC and the blue dot indicates the battery where the TOF was higher for a 0\% SoC)}
\label{fig3}
\end{figure}

A clear trend between the TOF estimation and the charged and discharged conditions was not obtained for every battery used in this study (Fig. \ref{fig3}). However, batteries 11, 21, and 33 (which are identified with a green dot in Fig. \ref{fig3}) showed an average higher TOF for the signals of each element on the 100\% SOC condition compared to the 0\% condition. Battery 13 showed a low standard deviation in its TOF estimations for all the elements compared to the other batteries. Meanwhile, battery 12 (identified with a blue dot in Fig. \ref{fig3}) had the opposite behavior where the full-discharge condition showed a higher TOF for the acoustic signal.

No significant difference between discharged and charged conditions was observed for the rest of the batteries as shown in Fig. \ref{fig3}.

\begin{figure}[htbp]
\centering
\centerline{\includegraphics[width=9cm]{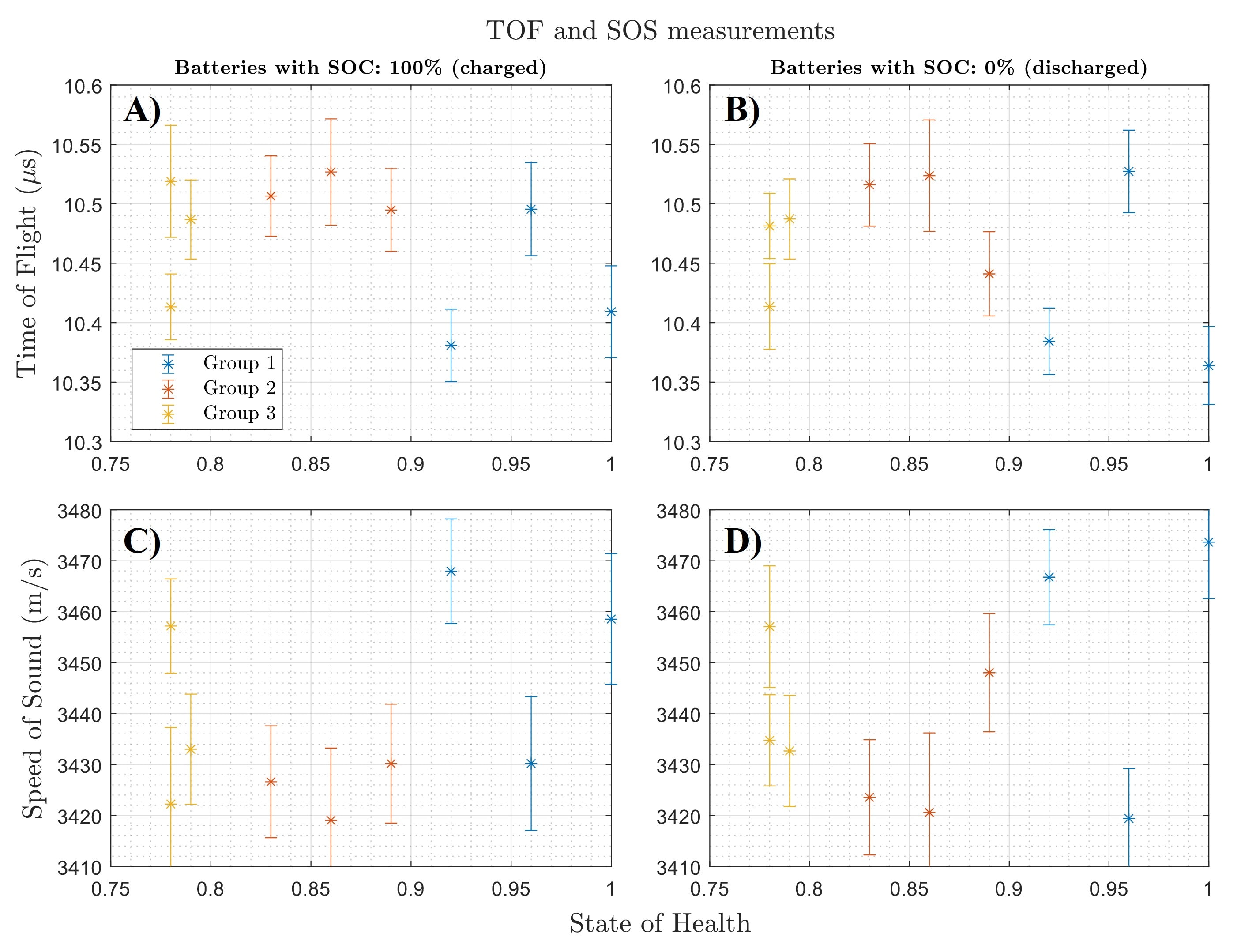}}
\caption{Time of flight measurements and speed of sound measurements against batteries’ state of health. \textbf{A)} TOF for full-charged condition, \textbf{B)} TOF for full-discharged condition, \textbf{C)} SOS for full-charged condition and \textbf{D)} SOS for full-discharged condition. Groups from Table \ref{tab1} are classified by colors}
\label{fig4}
\end{figure}

Likewise, there is not a clear trend between SOS/TOF with the SoH at this point as can be seen in Fig. \ref{fig4}.  However, for SoH between 0.93 and 0.86 an increase of TOF and a decrease of SOS was observed for higher levels of battery degradation, which means lower SoH values for both charged and discharged conditions.

\begin{figure}[htbp]
\centering
\centerline{\includegraphics[width=9cm]{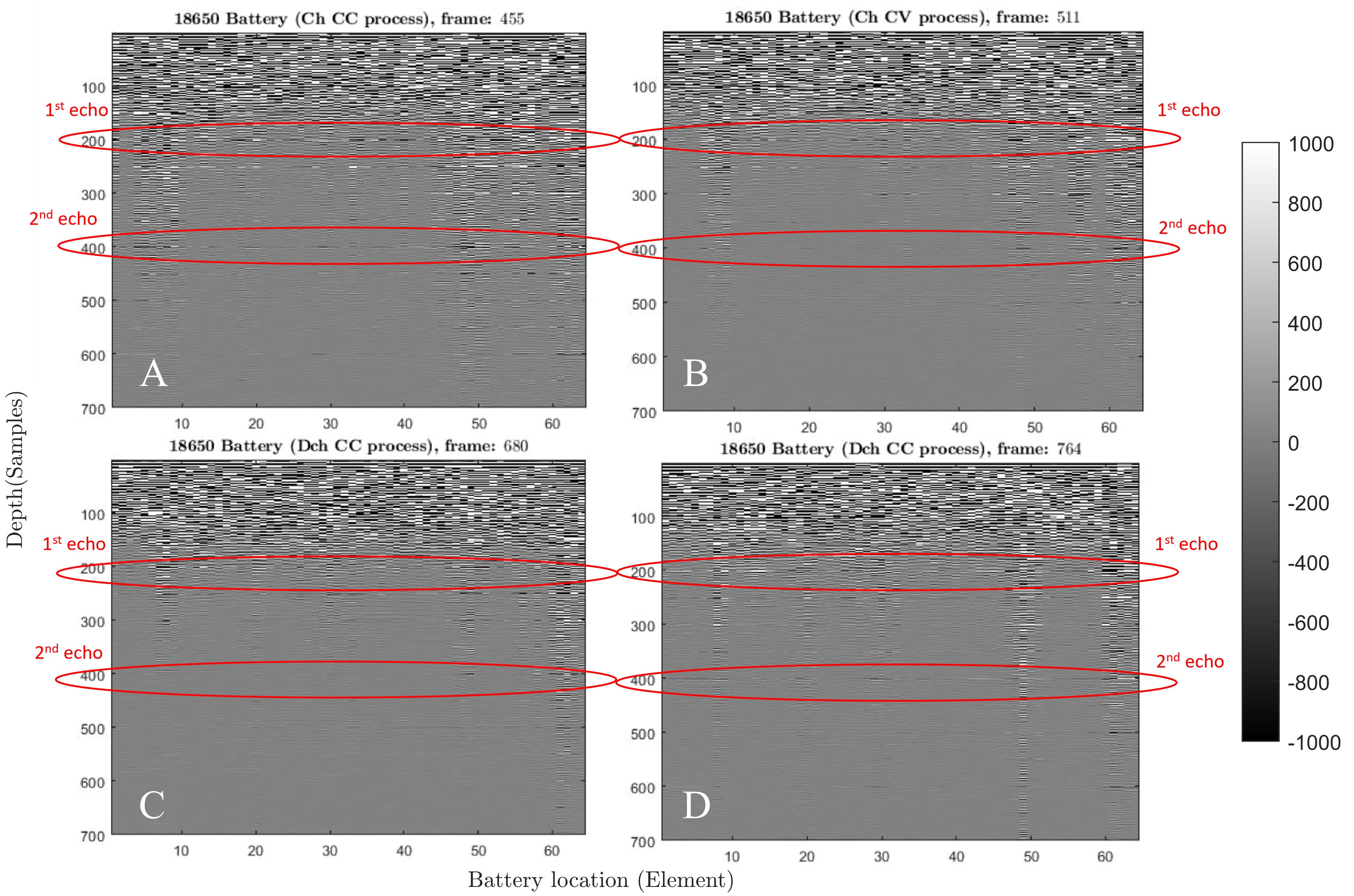}}
\caption{Ultrasonic response for the 64 channels for different stages of a charge/discharge cycle: \textbf{A)} start of the CC charge phase, \textbf{B)} start of the CV charge phase, \textbf{C)} start of the CC discharge stage, \textbf{D)} end of the process}
\label{fig5}
\end{figure}

The results of the “in operando” experiment are shown in Figure \ref{fig5}. In Panel A are the echoes obtained at the beginning of the CC phase of the charging process. Panel B shows the signals in the onset of the CV portion of the charge procedure. While Panels C and D show the starting of the CC discharge phase and the complete discharge state, respectively for the battery 13. No clear trends were observed for TOF and SOS estimations. However, a variation in the amplitude of the acoustic signal was identified during the charging/discharging process on the regions corresponding the echoes from the backwall (red areas in Fig. \ref{fig5}). In this case, the amplitude of the signals decreased during the charging and increased again in the course of the discharging.

\section{Discussion}

To the best of the authors’ knowledge, there are few works devoted to estimating the state of second-life batteries. These batteries have complex degradation pathways compare to factory-new cells or laboratory-aged batteries. In particular call coming from electric mobility applications tend to be subjected to abusive conditions depending on the users’ behavior and the environmental factors \cite{b14}. Even though the results from this work have not provide conclusive results, we are currently improving the measurement protocols and further experiments are being conducted to confirm these preliminary results. Nevertheless, this work showed that battery state has an influence on ultrasound measurements such as TOF and SOS, and could be used in the future for estimating the state of batteries. This approach is faster than conventional methods such as coulomb counting since it can be performed in very short time (ms) \cite{b15}.

Additionally, cylindrical batteries have a higher acoustic complexity compared to pouch cells, which are the most common reported cell geometry on the literature for ultrasound experiments. While pouch cells have a layered structure with clear differentiation between layers, the 18650-cell design typically consists of 15 to 25 layered windings \cite{b2}.

Furthermore, according to some authors, the two phases (CC and CV) during the charging process have contrary acoustic effects. For instance, in \cite{b17} is shown that there is a decrease in the acoustic TOF during the CC charging phase while there is an increase in the TOF during the CV section. Also, in the same study it was reported for the TOF an increase  in CC discharging and a decrease for relaxation periods where the battery is not forced with an electrical current. When comparing our results to this previous work, it is important to take into account that our measurements were taken in a SOC of 0\% and 100\%, therefore, we were not able to explore the changes inside both phases. Moreover, in our experiments the batteries had undergone the relaxation period described in \cite{b17}. 

The changes in amplitude in the “in operando” experiment suggest structural changes on each stage of the charge/discharge cycle, which lead to mechanical properties changes inside the battery and thus in the ultrasound waves propagation. However, the causes for these effects are difficult to explain at this time and would need further experiments and research.

\section{Conclusions and Future work}

In this study we used ultrasound measurements to explore the state of second-life batteries from an electric mobility commercial application. Even though there was no clear correlation between ultrasound battery response and battery state variables (such as SoC and SoH) at this point, we believe there is still a significant opportunity to use ultrasound to assess battery state in different applications. Further ’In operando’ experiments must be performed to track the changes in the ultrasound signals and their relationship with battery state variables.

As ultrasound wave characteristics such as TOF and SOS appear to be related to SoC and SoH, we believe that the analysis of ultrasound wave propagation inside Lithium-ion batteries will be a valuable tool to assess second-life applications and help in implementing a circular economy framework around batteries.

\section*{Acknowledgment}

This work was supported by the Royal Society of Engineering in the United Kingdom through its Transforming Systems through Partnership 20/21 call. Special acknowledgment  goes to Verasonics SAS and BATx for providing the ultrasound equipment and the second-life batteries used in this study respectively.


\begin{thebibliography}{00}
\bibitem{b1} S. Dühnen, J. Betz, M. Kolek, R. Schmuch, M. Winter, and T. Placke, “Toward Green Battery Cells: Perspective on Materials and Technologies,” Small Methods, vol. 4, no. 7, p. 2000039, Jul. 2020.

\bibitem{b2} G. Harper et al., “Recycling lithium-ion batteries from electric vehicles,” Nature, vol. 575, no. 7781, pp. 75–86, Nov. 2019  DOI:10.1038/s41586-019-1682-5ISBN:4158601916.
\bibitem{b3} S. I. Sun, A. J. Chipper, M. Kiaee, and R. G. A. Wills, “Effects of market dynamics on the time-evolving price of second-life electric vehicle batteries,” vol. 19, no. December 2017, pp. 41–51, 2018  DOI:10.1016/j.est.2018.06.012.
\bibitem{b4} M. Anna, F. Guarino, S. Longo, M. Ferraro, and M. Cellura, “Energy and environmental benefits of circular economy strategies: The case study of reusing used batteries from electric vehicles,” J. Energy Storage, vol. 25, no. April, p. 100845, 2019 DOI:10.1016/j.est.2019.100845.

\bibitem{b5} C. Pastor-Fernández, T. F. Yu, W. D. Widanage, and J. Marco, “Critical review of non-invasive diagnosis techniques for quantification of degradation modes in lithium-ion batteries,” Renew. Sustain. Energy Rev., vol. 109, pp. 138–159, Jul. 2019, doi: 10.1016/j.rser.2019.03.060.
\bibitem{b6} S. Montoya-Bedoya, L. Sabogal-Moncada, E. Garcia-Tamayo, and V. Martinez-Tejada, “A Circular Economy of Electrochemical Energy Storage Systems: Critical Review of SOH/RUL Estimation Methods for Second-Life Batteries,” in Intech, vol. i, 2020, p. 13.
\bibitem{b7} K. Laadjal and A. J. M. Cardoso, “Estimation of Lithium-Ion Batteries State-Condition in Electric Vehicle Applications: Issues and State of the Art,” Electronics, vol. 10, no. 13, p. 1588, Jun. 2021, doi: 10.3390/electronics10131588.
\bibitem{b8} X. Han et al., “A review on the key issues of the lithium ion battery degradation among the whole life cycle,” eTransportation, vol. 1, p. 100005, Aug. 2019, doi: 10.1016/j.etran.2019.100005.
\bibitem{b9} V. Sulzer et al., “The challenge and opportunity of battery lifetime prediction from field data,” Joule, vol. 5, no. 8, pp. 1934–1955, Aug. 2021, doi: 10.1016/j.joule.2021.06.005.
\bibitem{b10} Y. Wu, Y. Wang, W. K. C. Yung, and M. Pecht, “Ultrasonic Health Monitoring of Lithium-Ion Batteries,” Electronics, vol. 8, no. 7, p. 751, Jul. 2019, doi: 10.3390/electronics8070751.
\bibitem{b11} G. Davies et al., “State of Charge and State of Health Estimation Using Electrochemical Acoustic Time of Flight Analysis,” J. Electrochem. Soc., vol. 164, no. 12, pp. A2746–A2755, Sep. 2017, doi: 10.1149/2.1411712jes.
\bibitem{b12} A. G. Hsieh et al., “Electrochemical-acoustic time of flight: In operando correlation of physical dynamics with battery charge and health,” Energy Environ. Sci., vol. 8, no. 5, pp. 1569–1577, May 2015, doi: 10.1039/c5ee00111k.
\bibitem{b14} Guo, J.; Li, Y.; Pedersen, K.; Stroe, D.-I. Lithium-Ion Battery Operation, Degradation, and Aging Mechanism in Electric Vehicles: An Overview. Energies 2021, 14, 5220. https://doi.org/10.3390/en14175220
\bibitem{b15} Noura, N.; Boulon, L.; Jemeï, S. A Review of Battery State of Health Estimation Methods: Hybrid Electric Vehicle Challenges. World Electr. Veh. J. 2020, 11, 66. https://doi.org/10.3390/wevj11040066
\bibitem{b17} James B. Robinson, Martin Pham, Matt D.R. Kok, Thomas M.M. Heenan, Dan J.L. Brett, Paul R. Shearing, Examining the Cycling Behaviour of Li-Ion Batteries Using Ultrasonic Time-of-Flight Measurements, Journal of Power Sources, Volume 444, 2019, 227318, ISSN 0378-7753, https://doi.org/10.1016/j.jpowsour.2019.227318.


\end{thebibliography}
\end{document}